\documentclass[twocolumn,showpacs,preprintnumbers,superscriptaddress,prb,floatfix]{revtex4-1}

\usepackage{amsmath,amssymb}
\usepackage{graphicx,textcomp}
\usepackage{subfigure}

\newcommand{\figref}[1]{Fig.~\ref{#1}}
\renewcommand {\vec}    [1]    {\ensuremath{\mathbf{#1}}}
\newcommand   {\avg}    [1]    {\ensuremath{\left\langle#1\right\rangle}}
\newcommand   {\mat}    [1]    {\ensuremath{\mathbf{\bar{\bar{#1}}}} }			
\newcommand	  {\set}    [1]    {\ensuremath{ \{ #1 \} }}

\begin{document}

\title{Temperature dependent effective potential method for accurate free energy calculations of solids}
\author{Olle Hellman}
\affiliation{Department of Physics, Chemistry and Biology (IFM), Link\"oping University, SE-581 83, Link\"oping, Sweden.}

\author{Peter Steneteg}
\affiliation{Department of Physics, Chemistry and Biology (IFM), Link\"oping University, SE-581 83, Link\"oping, Sweden.}

\author{I. A. Abrikosov}
\affiliation{Department of Physics, Chemistry and Biology (IFM), Link\"oping University, SE-581 83, Link\"oping, Sweden.}

\author{S. I. Simak}
\affiliation{Department of Physics, Chemistry and Biology (IFM), Link\"oping University, SE-581 83, Link\"oping, Sweden.}

\begin{abstract}
We have developed a thorough and accurate method of determining anharmonic free energies, the temperature dependent effective potential technique (TDEP). It is based on \emph{ab initio} molecular dynamics followed by a mapping onto a model Hamiltonian that describes the lattice dynamics. The formalism and the numerical aspects of the technique are described in details. A number of practical examples are given, and results are presented, which confirm the usefulness of TDEP within \emph{ab initio} and classical molecular dynamics frameworks. In particular, we examine from first-principles the behavior of force constants upon the dynamical stabilization of body centered phase of Zr, and show that they become more localized. We also calculate phase diagram for $^4$He modeled with the Aziz \emph{et al.} potential and obtain results which are in favorable agreement both with respect to experiment and established techniques.
\end{abstract}

\maketitle

\section{Introduction}

One of the common goals for first principles calculations is the comparison of energies, such as configurational energies, surface energies, mixing enthalpies or lattice stabilities. Usually the comparison is limited to total energies. This is appropriate when the effects of temperature can be neglected. However, for many problems within physics, materials and Earth sciences they are not negligible and Gibbs free energy represents the proper thermodynamic potential when the temperature and pressure are external parameters. The quasiharmonic approximation can bridge the temperature gap, but there are cases where it falls short. Strongly anharmonic systems are not described well\cite{Grimvall2012}, especially dynamically unstable systems. Traditionally the problem of dynamical instability was addressed either by including more terms in the Taylor expansion of the potential energy or via a self-consistent approach.\cite{Klein1972,Born1951,Born1998}

Hooton\cite{Hooton} realised that even though the second derivatives at the equilibrium positions are negative, the atoms in a solid rarely occupy these positions. They move in the effective potential of their non-stationary neighbours. By sampling the potential energy surface not at the equilibrium positions but at the most probable positions for a given temperature one can obtain a harmonic approximation that describes the system at elevated temperatures. The self-consistent formalism employs an iterative procedure\cite{Gillis1968} by creating a harmonic potential, which is used to describe the thermal excitations that again give a new harmonic potential. This is then repeated until self-consistency.

The double-time Green's functions, developed by Choquard,\cite{choquard1967} use a cumulant expansion in the higher order terms. Although formally exact, this formalism requires knowledge of the higher order force constants. Obtaining these accurately for something other than the structurally simple systems is computationally very demanding from first principles.\cite{Chaput2011}

A recent implementation of the self-consistent formalism by Souvatzis \emph{et al.}\cite{Souvatzis2008a,Souvatzis2011a} uses \emph{ab initio} super\-cell calculations. A problem with this approach is that the excitations could only be done in the harmonic sense which means probing phase space with a limited basis set. Where the harmonic approximation works well this is not a problem. When the harmonic approximation fails it is due to a strong anharmonic contribution. Strongly anharmonic systems are by definition badly described with the harmonic approximation.

Several techniques use Born-Oppenheimer molecular dynamics to obtain anharmonic corrections to quasi-harmonic free energies.\cite{Grabowski2009c,Wu2009,Wu2010} They focus on anharmonic corrections to materials that are well described in the quasiharmonic approximation, and the applicability to strongly anharmonic systems is questionable when the phonon renormalization due to increased temperature can not be described by a linear equation.

We have developed a new method\cite{Hellman2011} that is similar to Hootons\cite{Hooton} original idea, but with a foundation in (ab initio) molecular dynamics. In this paper we present a substantial refinement and generalization of the temperature dependent effective potential method (TDEP), showing how it deals with: a model one-dimensional anharmonic oscillator, a strongly anharmonic system, bcc Zr, treated from first-principles, and $^4$He modelled with the Aziz \emph{et al.} potential.

\section{TDEP formalism}

The starting point of our method is to introduce a model Hamiltonian for a Bravais lattice in the harmonic form:
\begin{equation}\label{eq:harmhamiltonian}
\hat{H}=U_0+\sum_i \frac{\vec{p}_i^2}{2m_i}+\sum_{ij}\vec{u}_i\mat{\Phi}_{ij}\vec{u}_j
\end{equation}
which describes the lattice dynamics. Here $\vec{p}_i$ and $\vec{u}_i$ are the momentum and displacement of atom $i$, bold symbols indicate vectors and doubly overlined symbols matrices respectively. The reference point for the displacements is the 0K relaxed lattice (initially, in Sec. \ref{sec:intfree} we will revisit this). The interatomic force constants \mat{\Phi} and the ground state energy $U_0$ are yet to be determined. Given $N_a$ atoms in this model system the forces acting on the atoms are given by
\begin{equation}\label{eq:big_fdu}
\underbrace{\begin{pmatrix}
\vec{f}_1\\
\vec{f}_2 \\
\vdots \\
\vec{f}_{N_a} \\
\end{pmatrix}}_{\vec{F}^{\textrm{H}}_t}=
\underbrace{\begin{pmatrix}
\mat{\Phi}_{11} & \mat{\Phi}_{12} & \cdots & \mat{\Phi}_{1N_a} \\
\mat{\Phi}_{21} & \mat{\Phi}_{22} & \cdots & \mat{\Phi}_{2N_a} \\
\vdots & \vdots & \ddots & \vdots \\
\mat{\Phi}_{N_a1} & \mat{\Phi}_{N_a2} & \cdots & \mat{\Phi}_{N_aN_a} 
\end{pmatrix}}_{\mat{\Phi}}
\underbrace{\begin{pmatrix}
\vec{u}_1\\
\vec{u}_2 \\
\vdots \\
\vec{u}_{N_a} \\
\end{pmatrix}}_{\vec{U}_t}.
\end{equation}
As detailed in our previous paper\cite{Hellman2011} we seek to determine the force constant matrices through minimization of the difference in forces from the model system  a real system, simulated, for instance, by means of \emph{ab initio} molecular dynamics (AIMD). An AIMD simulation will provide a set of displacements $\set{\vec{U}_t^{\textrm{MD}}}$, forces $\set{\vec{F}_t^{\textrm{MD}}}$ and potential energies $\set{E_t^{\textrm{MD}}}$. We seek to minimize the difference in forces from AIMD and our harmonic form ($\vec{F}^{\textrm{H}}$) at time step $t$, summed over all time steps $N_t$:
\begin{equation}\label{eq:min_f}
\begin{split}
\min_{\mat{\Phi}}\Delta \vec{F} & =\frac{1}{N_t} \sum_{t=1}^{N_t}  \left| \vec{F}_t^{\textrm{MD}}-\vec{F}_t^{\textrm{H}} \right|^2= \\
& =\frac{1}{N_t} \sum_{t=1}^{N_t} \left| \vec{F}_t^{\textrm{MD}}-\mat{\Phi}\vec{U}_t^{\textrm{MD}} \right|^2 = \\
& = \frac{1}{N_t} \left\Vert 
\begin{pmatrix} \vec{F}_1^{\textrm{MD}} \ldots \vec{F}_{N_t}^{\textrm{MD}} \end{pmatrix}-\mat{\Phi}
\begin{pmatrix} \vec{U}_1^{\textrm{MD}} \ldots \vec{U}_{N_t}^{\textrm{MD}} \end{pmatrix}
\right\Vert.
\end{split}
\end{equation}
This is realized with a a Moore-Penrose pseudoinverse
\begin{equation}\label{eq:solve_for_tilde_D}
\mat{\Phi}=\begin{pmatrix} \vec{F}_1^{\textrm{MD}} \ldots \vec{F}_{N_t}^{\textrm{MD}} \end{pmatrix}
\begin{pmatrix} \vec{U}_1^{\textrm{MD}} \ldots \vec{U}_{N_t}^{\textrm{MD}} \end{pmatrix}^{+}
\end{equation}
to obtain the linear least squares solution for $\mat{\Phi}$ that minimize $\Delta \vec{F}$. This is then mapped to the form
\begin{equation}\label{eq:starform}
\mat{\Phi} \longrightarrow \Phi^{\alpha\beta}_{\mu\nu}(\vec{R}_{l}),
\end{equation}
where $\alpha\beta$ are indices to atoms in a unit cell with $N_{uc}\le N_a$ atoms and $\mu\nu$ Cartesian indices. The pair vectors in the supercell $\vec{R}_{ij}$ are mapped to stars of lattice vectors $\vec{R}_l$ connecting atoms of type $\alpha$ and $\beta$. From this form the phonon dispersion relations, free energy and all other quantities can be extracted. This direct implementation works well,\cite{Hellman2011} but the numerical efficiency can be improved, as is demonstrated below.

\section{Symmetry constrained force constant extraction}

The form of the force constant matrices depends only on the supercell used in the AIMD and the crystal lattice. We begin by populating the force constant matrices $\Phi^{\alpha\beta}_{\mu\nu}(\vec{R}_{l})$ with unknown variables $\theta_k$
\begin{equation}
\begin{split}
\mat{\Phi}^{11}(\vec{R}_{1})=& \begin{pmatrix}
\theta_1 & \theta_2 & \theta_3 \\
\theta_4 & \theta_5 & \theta_6 \\
\theta_7 & \theta_8 & \theta_9 \\
\end{pmatrix}\\
\mat{\Phi}^{11}(\vec{R}_{2})=& \begin{pmatrix}
\theta_{10} & \theta_{11} & \theta_{12} \\
\theta_{13} & \theta_{14} & \theta_{15} \\
\theta_{16} & \theta_{17} & \theta_{18} \\
\end{pmatrix}
\end{split}
\end{equation}
and so on, including vectors $\vec{R}_l$ up to a cutoff determined by the size of the supercell. Some of these $\theta_k$ are equivalent. To find out which, we look at the symmetry relations the force constant matrices have to obey. We have\cite{Maradudin1968}
\begin{align}
\label{eq:fcm_s_1}\sum_{l\alpha} \mat{\Phi}^{\alpha\beta}(\vec{R}_{l})&=0 \hspace{2mm}\textrm{for each } \beta \\
\label{eq:fcm_s_2}\mat{\Phi}^{\alpha\beta}(\vec{R}_{l}) &=\mat{\Phi}^{\beta\alpha}(-\vec{R}_{l}) \\
\label{eq:fcm_s_3}\Phi^{\alpha\beta}_{\mu\nu}(\vec{R}_{l})&=\Phi^{\alpha\beta}_{\nu\mu}(\vec{R}_{l})
\end{align}
that stem, in order, from the facts that there is no net translation of the crystal, all Bravais lattices have inversion symmetry, and that the order of the second derivatives does not matter. Each relation will give us a few equations for the unknowns $\theta_k$, reducing their number. In addition to these fundamental properties of the force constant matrices we can benefit from the symmetry of the lattice. If symmetry relation $S$, belonging to the point group of the lattice, relates two vectors $\vec{R}_l=S\vec{R}_k$ we have the following relation:
\begin{equation}\label{eq:fcm_symops}
\mat{\Phi}^{\alpha\beta}(\vec{R}_{l}) =S\mat{\Phi}^{\alpha\beta}( \vec{R}_{k} ) S^{T}
\end{equation}
By applying equations \eqref{eq:fcm_s_1}--\eqref{eq:fcm_symops} the number of unknown variables is massively reduced. For example, a bcc lattice modelled as a $4\times4\times4$ supercell (128 atoms) would have 147456 unknown variables in $\mat{\Phi}$, if one does not consider symmetry arguments. Application of Eqs. \eqref{eq:fcm_s_1}--\eqref{eq:fcm_symops} reduces the problem to 11 unknown variables. Having found the reduced problem with $N_\theta$ unknown variables, it can be substituted back into
\eqref{eq:big_fdu}. The expression for the forces at timestep $t$ will schematically look like this:
\begin{equation}\label{eq:big_fdu_phi}
\underbrace{\begin{pmatrix}
f_1\\
f_2 \\
\vdots \\
f_\gamma \\
\vdots \\
f_{3N_a} \\
\end{pmatrix}}_{\vec{F}_t^{\textrm{H}}}=
\underbrace{\begin{pmatrix}
\theta_1 & 0 & 0 &  \ldots \\
0 & \theta_1 & 0 &  \ldots \\
0 & 0 & \theta_1 &  \ldots \\
\theta_2 & \theta_3 & -\theta_4 & 	\ldots \\
\theta_3 & -\theta_2 & 0 	&	\ldots \\
-\theta_4 & 0 & 0 	&		\ldots \\
\vdots & \vdots & \vdots &  \ddots
\end{pmatrix}}_{\mat{\Phi}}
\underbrace{\begin{pmatrix}
u_1 \\
u_2 \\
\vdots \\
u_\delta \\
\vdots \\
u_{3N_a} \\
\end{pmatrix}}_{\vec{U}_t}.
\end{equation}
The actual distribution of the $\theta_k$ will depend on the problem at hand. Carrying out the matrix product gives us a new set of equations for the forces
\begin{equation}
f_\gamma=\sum_k \theta_k \sum_{\delta}c^k_{\gamma\delta} u_\delta.
\end{equation}
where second sum describes the coefficients for each $\theta_k$ contained in the expression for force component $f_\gamma$. These coefficients are linear combinations of the displacement components $u_\delta$. The explicit form is determined by the lattice. In matrix form this is written as
\begin{equation}\label{eq:FeqCPhi}
\vec{F}=\mat{C}(\vec{U})\vec{\Theta} , \quad C(\vec{U})_{k\gamma}=\sum_{\delta} c^k_{\gamma\delta}u_\delta.
\end{equation}
Eq. \ref{eq:FeqCPhi} is equivalent to Eq. \ref{eq:big_fdu}, it is just rewritten in terms of the symmetry inequivalent interactions. This implementation symbolically reduces the number of unknowns, generates the function that gives the matrix $\mat{C}$ from a set of displacements $\vec{U}$ and the mapping from the set of $\theta_k$ back to the force constant matrix $\Phi^{\alpha\beta}_{\mu\nu}(\vec{R}_{l})$.\footnote{In practice, this is implemented in Mathematica} 

To find $\Theta$ we minimize the difference in forces between AIMD simulations and the model hamiltonian
\begin{equation}\label{eq:min_f_phi}
\begin{split}
\min_{\Theta}\Delta \vec{F} & =\frac{1}{N_t} \sum_{t=1}^{N_t}  \left| \vec{F}_t^{\textrm{MD}}-\vec{F}_t^{\textrm{H}} \right|^2= \\
& =\frac{1}{N_t} \sum_{t=1}^{N_t} \left| \vec{F}_t^{\textrm{MD}}-\mat{C}(\vec{U}_t^{\textrm{MD}})\vec{\Theta} \right|^2 = \\
& = \frac{1}{N_t} \left\Vert 
\begin{pmatrix} \vec{F}_1^{\textrm{MD}} \\  \vdots \\ \vec{F}_{N_t}^{\textrm{MD}} \end{pmatrix}-
\begin{pmatrix} \mat{C}(\vec{U}_1^{\textrm{MD}}) \\ \vdots \\ \mat{C}(\vec{U}_{N_t}^{\textrm{MD}}) \end{pmatrix}\vec{\Theta}
\right\Vert.
\end{split}
\end{equation}
This is again realised with the least squares solution for $\vec{\Theta}$ 
\begin{equation}\label{eq:phi_eqFC}
\vec{\Theta}=
\begin{pmatrix}
\mat{C}(\vec{U}_1^{\textrm{MD}}) \\
\vdots \\
\mat{C}(\vec{U}_{N_t}^{\textrm{MD}})
\end{pmatrix}^{+}
\begin{pmatrix}
\vec{F}^{\textrm{MD}}_1 \\
\vdots \\
\vec{F}^{\textrm{MD}}_{N_t}
\end{pmatrix}.
\end{equation}
Having determined $\vec{\Theta}$ we can substitute the components into the force constant matrix and proceed to calculate thermodynamic properties of the original (real) system.

The suggested scheme is a superior way of using symmetry to improve the numerical accuracy. Most schemes involving symmetry revolve around deter\-mining the interaction in one direction and then using the symmetry relations to translate and rotate that interaction. Any numeri\-cal errors--which are always present--will propagate to all interactions, whereas in the present approach the errors will be averaged and should cancel each other to a large degree. 

The advantage of this procedure will be illustrated below.


\begin{figure}[htb]
\includegraphics[width=\linewidth]{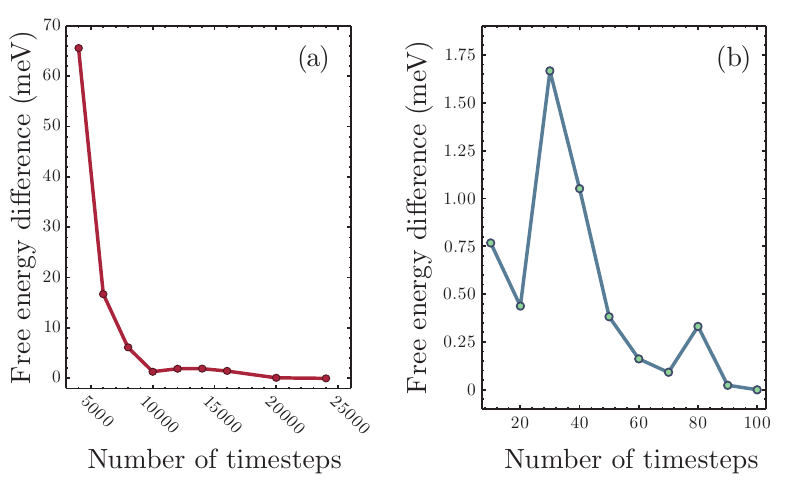}
\caption{\label{fig:konv1}(color online) Convergence of the free energy of bcc Zr at 1300K with respect to the number of timesteps. In panel (a) symmetry is treated numerically and in (b) it is treated in the novel analytical way. The same input data is used in both cases, and it converges to the same value, but we have improved the performance by several orders of magnitude.}
\end{figure}

\section{Internal degrees of freedom}\label{sec:intfree}

If the system has internal degrees of freedom for the structural relaxations, such as a crystal with point defects, an interface or a random alloy the atoms ideal positions and equilibrium positions do not coincide. While one could find the relaxed positions from 0K calculations, the equilibrium positions are by no means constant with respect to temperature. TDEP handles this in an elegant way. Note that we find the second order terms in \eqref{eq:harmhamiltonian} with a least squares fit of the slope of force versus displacement. Originally, the displacements could be calculated with respect to the wrong equilibrium positions that do not correspond to the temperature of the simulation. Still our experience shows that slope will be the well approximated. That allows for the following procedure:
\begin{itemize}
	\item[--] Guess equilibrium positions, usually the ideal lattice positions.
	\item[--] Use these to calculate the displacements $u$ from the AIMD simulations.
	\item[--] Determine $\vec{\Theta}$ and from that the residual force
	\begin{equation}
	\vec{F}_{r}=\sum_{t=1}^{N_t} \frac{1}{N_t} \left( \vec{F}_t^{\textrm{MD}} - \vec{F}_t^{\textrm{H}}\right),
	\end{equation}
	where $\vec{F}_t^{\textrm{MD}}$ are the AIMD forces and $\vec{F}_t^{\textrm{H}}$ are given by equation \ref{eq:FeqCPhi}. $N_t$ is the number of timesteps, and subscript $t$ denote the corresponding forces at time step $t$.
	\item[--] These forces are then used to move the atoms in a steepest descent scheme towards equilibrium positions. The whole process is repeated until convergence.
\end{itemize}

Our test shows that this procedure is remarkably stable. The second order force constants $\Phi$ are weakly affected by the choice of equilibrium positions. The vibrational entropy and phonon dispersion relations are largely unaffected as well. Eliminating the first order term, however, is formally important and crucial when extracting higher order terms. Note that self-consistent iterations are numerically efficient, because the most time-consuming step for applications of TDEP is MD simulations, while their mapping on model Hamiltonian \eqref{eq:harmhamiltonian} represents post-processing of the MD results with minimal computaional cost.

\section{Determining the free energy}\label{sec:freeen}

We will begin by reiterating the traditional way how free energy is determined in the quasiharmonic approximation. Divided into parts it will be
\begin{equation}
\begin{split}
F & = U-TS \\
& = \underbrace{U_{\textrm{tot}}-TS_{el}}_{F_{\textrm{el}}}+\underbrace{\avg{E_k}+\avg{U_{\textrm{vib}}}+U_{\textrm{zp}}-TS_{\textrm{vib}}}_{F_{\textrm{vib}}},
\end{split}
\end{equation}
where a division of the right hand side into parts makes a clear distinction between the electronic contribution $F_{\textrm{el}}$ and the vibrational contribution $F_{\textrm{vib}}$. The electronic contribution is divided into the total energy of the lattice, $U_{\textrm{tot}}$, and the electronic entropy $S_{el}$. The vibrational contribution is divided into average kinetic $E_k$ and potential energy $U_{\textrm{vib}}$ of the ions, vibrational entropy $S_{\textrm{vib}}$ and zero point energy $U_{\textrm{zp}}$. The lattice contribution is obtained from DFT calculations with the Mermin functional and the vibrational part from the harmonic approximation via
\begin{equation}\label{eq:phonon_free_energy}
F_{\textrm{vib}}= \int_0^\infty g(\omega)\left[ k_B T
\ln \left( 1- \exp \left( -\frac{\hbar\omega}{k_B T} \right) \right)+ \frac{\hbar \omega}{2}\right] d\omega,
\end{equation}
Where $g(\omega)$ is the phonon density of states. In this approach, all the vibrational contributions are calculated within the harmonic approximation. 

Turning to AIMD, the free energy (in the canonical ensemble) is divided as
\begin{equation}
F=\avg{U_{\textrm{MD}}}+\avg{E_k}-TS_{\textrm{MD}},
\end{equation}
were the potential energy $U_{\textrm{MD}}$ is temperature dependent. Since the ions move as classical particles the zero point energy is missing. There is unfortunately no information about the entropy, but through the force constant matrices obtained using TDEP, the vibrational entropy and zero point energy can be estimated. For TDEP to have an accurate free energy the potential energy should, on average, be equal to that of AIMD. This would ensure that the full anharmonic $\avg{U_{\textrm{MD}}}$ is included. The problem is that $U_{\textrm{MD}}$ is rapidly oscillating over time, requiring a long simulation time to converge. If we look at the TDEP potential energy
\begin{equation}\label{eq:harmenagain}
U_{\textrm{TDEP}}(t)= U_0+\frac{1}{2} \sum_{ij\alpha\beta} 
\Phi^{\alpha\beta}_{ij} u^\alpha_{i}(t) u^\beta_{j}(t)
\end{equation}
and recognise that it should model the thermal fluctuations of the original system we can overcome the numerical issues. Setting the average potential energies equal, $\avg{U_{\textrm{MD}}}=\avg{U_{\textrm{TDEP}}}$, gives us
\begin{equation}\label{eq:unoll_from_md}
U_0=\avg{U_{\textrm{MD}}(t)-\sum_{ij\alpha\beta}\frac{1}{2} 
\Phi^{\alpha\beta}_{ij} u^\alpha_{i}(t) u^\beta_{j}(t)}
\end{equation}
By removing the thermal excitations of the harmonic form the fluctuations can be decreased by roughly an order of magnitude and the accuracy of $U_0$ is thus increased be the same amount. Including higher order terms in the energy expansion would further serve to minimize these fluctuations.

\begin{figure}
\includegraphics[width=\linewidth]{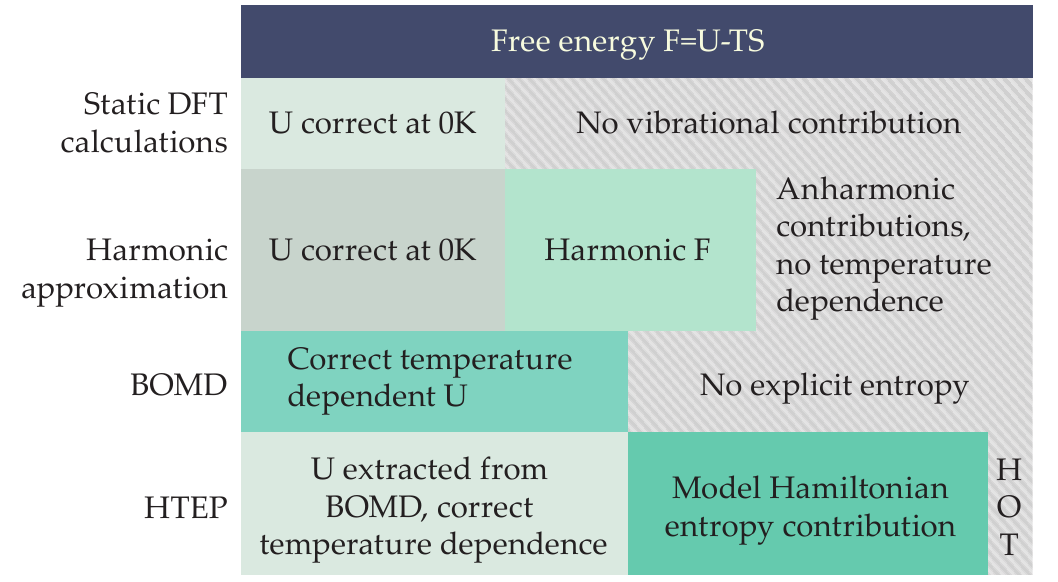}
\caption{\label{fig:division_of_free_energy}%
Illustration of the different terms included in free energy calculations using different approaches. The filled boxes denote the terms that are included, and the striped areas what is omitted. The main message is to point out that the internal energy has a non-trivial temperature dependence, something that is omitted in the quasiharmonic approximation. HOT indicate the higher order terms that are missing in the TDEP free energy, Eq. \ref{eq:tdep_helmholtz} and \ref{eq:phonon_free_energy}.}
\end{figure}

The potential energy that was removed will be added again when the Helmholtz free energy is calculated:
\begin{equation}\label{eq:tdep_helmholtz}
F_{\mathrm{TDEP}}=U_0+F_{\mathrm{vib}},
\end{equation}
where $F_{\textrm{vib}}$ is the phonon contribution given Eq. \ref{eq:phonon_free_energy}, with the phonon density of states determined in the TDEP formalism. It includes the kinetic and potential energy of the ions. In \figref{fig:division_of_free_energy} we further illustrate the difference in methods of obtaining the free energy.

The formally exact method of thermodynamic integration\cite{frenkel2002understanding} can be used to determine the free energy. This method will determine the anharmonic correction to the free energy. If the TDEP model Hamiltonian is used as the reference point the full free energy will be 
\begin{equation}\label{eq:termoint}
F=U_0+F_{\mathrm{vib}}+\underbrace{\int_0^1 \avg{U_{\textrm{MD}}-U_{\textrm{TDEP}}}_\lambda d\lambda}_{\Delta F_{\textrm{AH}}}.
\end{equation}
The integral is over the Kirkwood coupling parameter $\lambda$, and the potential energy difference is between the model Hamiltonian and the molecular dynamics potential energy. 

The model TDEP Hamiltonian is constructed to describe the system as accurately as possible while retaining the harmonic form. It is then easy to argue that the anharmonic correction term $\Delta F_{\textrm{AH}}$ should be small. While it is difficult to make general statements regarding this, our experience so far is that this correction is very small in every system we have tested.

\begin{figure}[htb]
\includegraphics[width=\linewidth]{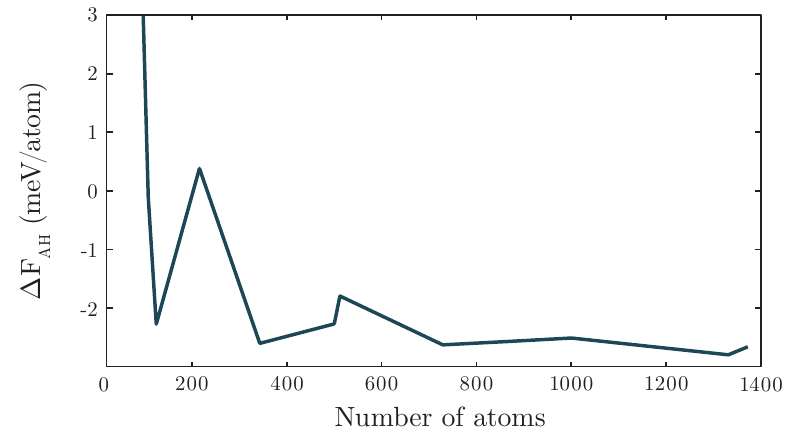}
\caption{\label{fig:konv2}(color online) Convergence of the free energy correction from thermodynamic integration ($\Delta F_{AH}$ in Eq. \ref{eq:termoint}) with respect to system size. At system sizes smaller than $\sim$700 atoms the uncertainty is of about the same order as the correction. This particular case is for fcc Cu modelled with an embedded atom potential\cite{Finnis1984}.}
\end{figure}

In addition, the thermodynamic integration technique can be numerically inefficient when high accuracy is needed.  While one can accurately control the numerical accuracy\cite{Grabowski2009c} the finite size effects are more difficult to control, especially in \emph{ab intio} simulations. In \figref{fig:konv2} we show that the error due to the limited size is on the same order as the correction to the TDEP free energy in reasonable simulation sizes%
\footnote{The thermodynamic integrations were carried out from a TDEP potential extracted for fcc Cu modelled with an embedded atom potential\cite{Finnis1984}. A Langevin thermostat was used to control the temperature and break the mode-locking. The numerical integration over coupling parameter $\lambda$ was carried out over 15 discrete steps, and the we ran the MD simulations for $\sim$ 300000 timesteps to ensure convergence within 0.1meV/atom.}%
. It makes little use to add a correction to the TDEP free energy where the uncertainty is of the same order of magnitude as the correction itself. 

\begin{figure}[htb]
\includegraphics[width=\linewidth]{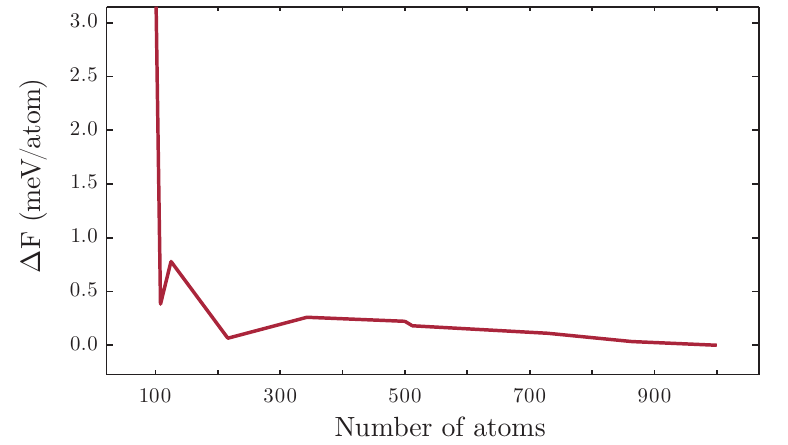}
\caption{\label{fig:konv3}(color online) Convergence of the free energy from TDEP with respect to simulation size. The system in question is fcc Cu with a classical embedded atom potential\cite{Finnis1984}. At sizes about 100 atoms it is converged within 1meV/atom.}
\end{figure}

The TDEP free energy, on the other hand, behaves well with respect to limited simulation cell. In figure \ref{fig:konv3} we see that at the reasonable system size of 100 atoms the free energy is converged within 1meV/atom.%
\footnote{The convergenve tests for TDEP used an an embedded atom potential\cite{Finnis1984} for fcc Co. We used A Langevin thermostat and ran the MD simulations for $\sim$ 150000 timesteps to ensure convergence within 0.1meV/atom.}
It is also easily converged in terms of simulation length: in \figref{fig:konv1} we illustrate the advantage of analytically treating symmetry. In our previous work,\cite{Hellman2011} we studied Zr in the bcc phase. There, we found convergence within 1meV/atom for the free energies after 25000 time steps. With the symmetry constraints, we converged to the same value using 50 time steps, an improvement by several orders of magnitude.

\begin{figure}[htb]
\includegraphics[width=\linewidth]{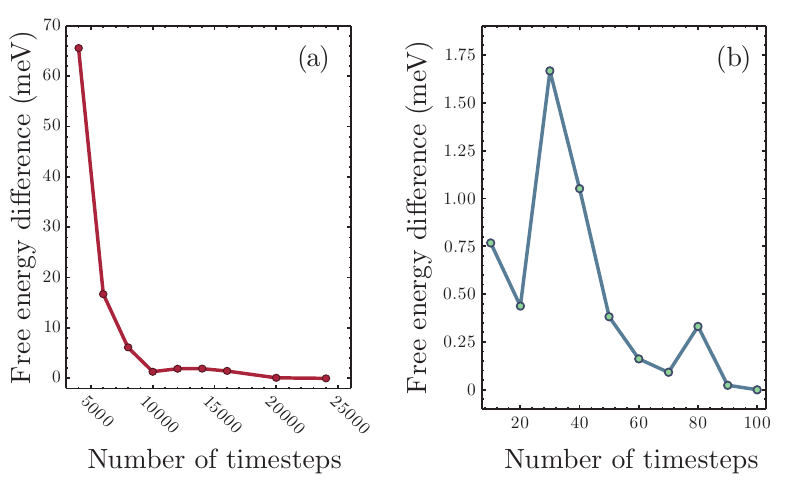}
\caption{\label{fig:konv1}(color online) Convergence of the free energy of bcc Zr at 1300K with respect to the number of timesteps. In panel (a) symmetry is treated numerically and in (b) it is treated in the novel analytical way. The same input data is used in both cases, and it converges to the same value, but we have improved the performance by several orders of magnitude.}
\end{figure}

\section{Application of TDEP to a one-dimensional anharmonic oscillator}

\begin{figure}[htb]
\includegraphics[width=\linewidth]{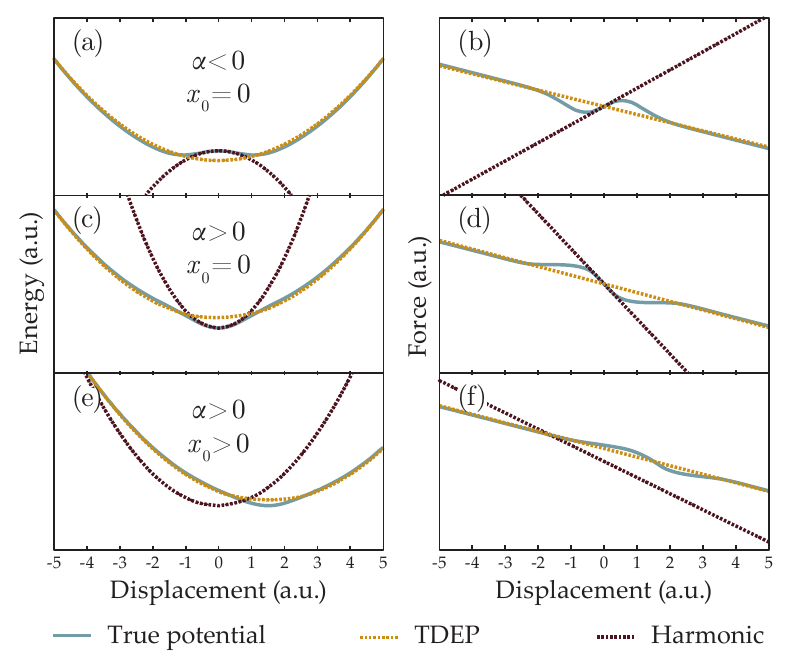}
\caption{\label{fig:anharmosc}Comparison of performance of TDEP and the harmonic Taylor expansion for the potential described by equation \ref{eq:almostharmonic}. Three examples are shown when the conventional harmonic approximation fails to describe the potential while TDEP succeeds. Panels (a),(c) and (e) show the potentials and (b), (d) and (f) show the forces. $\alpha$ and $x_0$ are the parameters of Eq. \ref{eq:almostharmonic}. In panel (a) the harmonic approximation correspongd to a dynamically unstable system, whereas TDEP provides a dynamically stable solution. In panel (b) the harmonic approximation provides an inaccurate potential. Panel (c) shows how TDEP finds the high temperature equilibrium position $x_0$.}
\end{figure}

To illustrate TDEP we first apply it to a one-dimensional anharmonic oscillator. Consider the following one-dimensional potential:
\begin{equation}\label{eq:almostharmonic}
U(x)=\frac{k(x-x_0)^2}{2}+\alpha e^{-\beta(x-x_0)^2}.
\end{equation}
Here $x$ is the position and $k$, $\alpha$ and $\beta$ are known parameters. The equilibrium position can depend on temperature and is assumed to be 0 at $T=0K$. The aim is to find the second degree polynomial fit to Eq.~\ref{eq:almostharmonic} that best describes the system. If this polynomial only consist of a quadratic and a constant term it will describe a harmonic oscillator with well defined free energy. If one applies the harmonic approximation to this potential it will not work well. The second derivative
\begin{equation}
\frac{d^2U}{dx^2}=k-2\alpha\beta e^{-\beta(x-x_0)^2}+(4\alpha\beta(x-x_0))^2e^{-\beta(x-x_0)^2}
\end{equation}
will determine the force constant $\Phi$. The temperature dependence of $x_0$ is omitted and we will end up with
\begin{equation}
\Phi=\left. \frac{d^2U}{dx^2} \right|_{x=0}=k+2\alpha\beta(2\alpha\beta x_0-1)e^{-\beta x_0^2}.
\end{equation}
This, as seen in \figref{fig:anharmosc}, will not be a particularly good model for the true potential. These issues arise from the fact that the potential energy surface is only probed at $x=0$, the $T=0$ equilibrium positions. To work around this problem, let us apply TDEP and put a particle in the potential given by equation \eqref{eq:almostharmonic} and perform a MD simulation. When controlled by an appropriate thermostat, the particle will yield a set of $N_t$ forces, positions and energies, \set{F_t,x_t,E_t}, one for each timestep. This data can now be used to fit a potential of the form
\begin{figure}[htb]
\includegraphics[width=\linewidth]{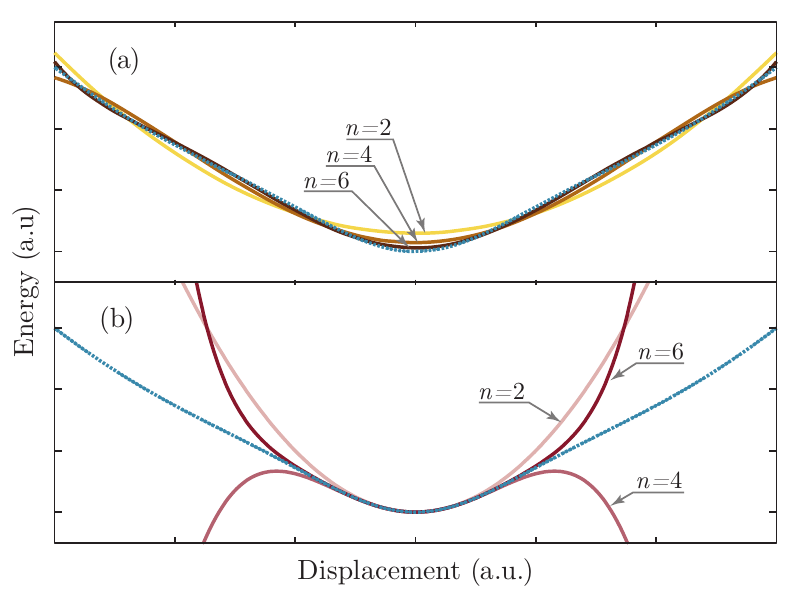}
\caption{\label{fig:highorder}Comparison of the effects of including higher order terms between the harmonic approximation and TDEP. When extending the Taylor expansion of an anharmonic potential (dashed blue line) in the Born--von Karman ansatz to higher order terms we end up with the series of lines depicted in panel (b). Panel (a) shows the same extension for TDEP. Even limiting oneself to the second order term ($n=2$), the fit will implicitly contain anharmonism to an arbitrary degree in the range that is thermally accessible. Extending TDEP to a higher order converges towards the true potential faster than when higher order terms are added to the harmonic approximation.}
\end{figure}
\begin{equation}\label{eq:almostharmonichamiltonian}
U(x)=\tilde{\Phi}^{(0)}+\frac{1}{2}\tilde{\Phi}^{(2)} (x-x'_0)^2,
\end{equation}
a harmonic potential centered at $x'_0$. Let us begin by determining the second order term. As discussed in Sec. \ref{sec:intfree} we guess a value for $x'_0$, use the forces from molecular dynamics \set{F_t} and minimize
\begin{equation}\label{eq:llsqfit1d}
\Delta F = \frac{1}{N_t}\sum_{t=1}^{N_t}| F_t - \tilde{\Phi}^{(2)} (x_t-x'_0)|.
\end{equation}
This is easiest realized as a least squares fit of a straight line in forces, as demonstrated in the right panels of \figref{fig:anharmosc}. Equation \eqref{eq:llsqfit1d} determines the second order term. The residual force at $x'_0$, $\Delta F$, can be used to find the equilibrium position. It is done in the following manner: a guess for $x'_0$ gives us a $\tilde{\Phi}^{(2)}$ and $\Delta F$. This residual force is used to move $x'_0$ to a new position, and the process is repeated until self-consistency is reached. When we have found the equilibrium position we can safely assume that any first order term in our polynomial can be set to 0. As described in Sec. \ref{sec:freeen} The constant energy term, $\Phi^{(0)}$, can be determined from the energies $\set{E_t}$ obtained from molecular dynamics simulations:
\begin{equation}
\Phi^{(0)}=\avg{E_t-\frac{1}{2}\tilde{\Phi}^{(2)} (x_t-x'_0)^2}_t.
\end{equation}
This is the best possible potential of the harmonic form at a given temperature that approximates the original potential in Eq. \ref{eq:almostharmonic}. In \figref{fig:anharmosc} the true potential and the fit are illustrated for different $\alpha$,$\beta$ and $x_0$. The anharmonism of the potential is implicitly described by the polynomial fit. In \figref{fig:highorder} the expansion in Eq. \ref{eq:almostharmonichamiltonian} has been extended to higher orders for an anharmonic potential. TDEP, probing the effective potential at finite temperature, converges to the true potential rapidly whereas including more terms in the Taylor expansion in Eq. \eqref{eq:harmhamiltonian} does by no means guarantee numerical stability at finite temperature.

\section{Practical applications of TDEP}

Let us summarize this and present the scheme used to calculate accurate Gibbs free energy surfaces in the TDEP formalism from first principles.

\begin{flushleft}

\begin{itemize}
\item[--] First calculate DFT total energy as a function of volume. This provides an equation of state and allows us to choose the volume interval, which covers pressure of interest.

\item[--] If feasible, obtain approximate harmonic potentials for the systems at hand in this volume interval. These potentials are used to speed up the calculations as described in Steneteg \emph{et al.} (submitted to PrB, BZ11707).

\item[--] On the grid of volumes and temperatures, perform AIMD simulations in the canonical ensemble.

\item[--] From these simulations, extract internal energy $U_0$ \eqref{eq:unoll_from_md} and interatomic force constants using Eq. \ref{eq:min_f_phi}, ensuring convergence of the free energy with respect to simulation length.

\item[--] To increase further the accuracy of the calculation, it is recommended to select a subset of uncorrelated samples from the AIMD simulations upsample these to high accuracy, as described in Ref.~\onlinecite{Grabowski2009c}. A new free energy is calculated.

\item[--] The equation of state is interpolated over the grid of temperatures and volumes providing the Gibbs free energy surface. This is then repeated for each structure, compound or composition of interest.
\end{itemize}
\end{flushleft}

\begin{figure}[h!]
\includegraphics[width=\linewidth]{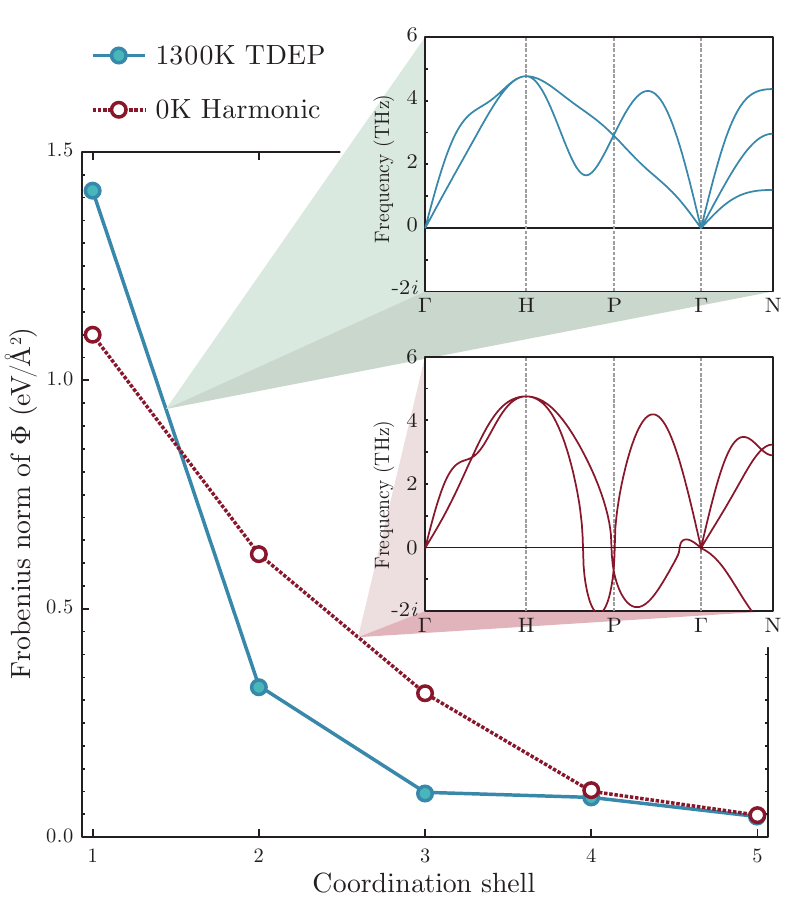}
\caption{\label{fig:frobnormar}(color online) Comparison of the TDEP and quasiharmonic force constant matrices. We have plotted the Frobenius norm of the force constants versus coordination shell. In the inset we show the corresponding phonon dispersion relations. The empty circles are 0K harmonic values and the filled circles are TDEP values extracted at 1300K. At high temperature the interactions at close distances are stronger, and fall off faster with increasing distance.}
\end{figure}

\begin{figure}
\includegraphics[width=\linewidth]{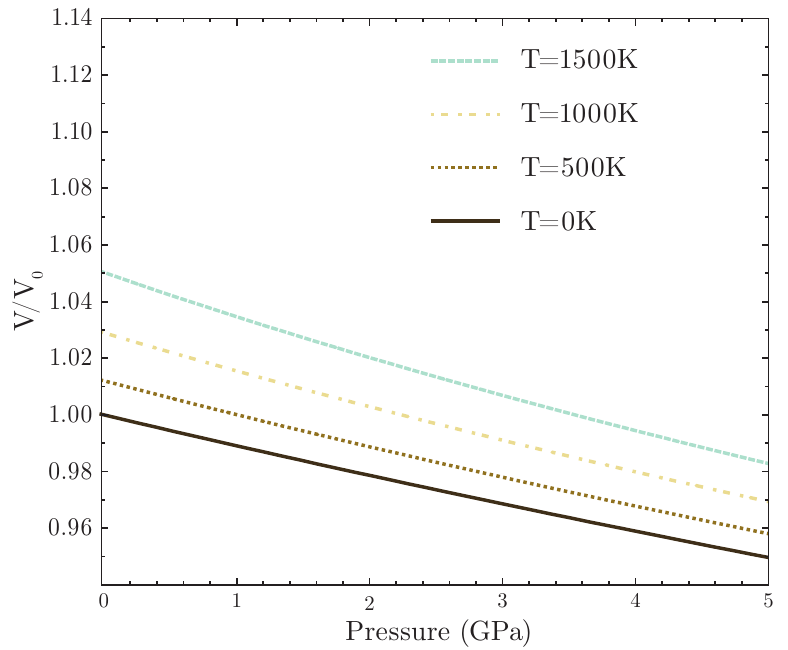}
\caption{\label{fig:zr_eos}(color online) Equation of state for bcc Zr calculated with TDEP method at temperatures 500K (dashed line), 1000K dot-dashed line, and 1500K (dotted line). Equation of state obtained by means of conventional DFT calculations at T=0K  is also shown with full line, and the zero temperature equilibrium volume V$_0$=22.83\AA$^3$ is chosen as a reference point at all the temperatures. Note, that bcc Zr is dynamically unstable at T=0K, see the bottom inset in Fig. 7.}
\end{figure}

TDEP is a thorough and time-consuming method, but the results are excellent. The phonon dispersion relations of a material that is dynamically unstable at zero temperature is a good example. When re-evaluating the results for Zr obtained in Ref. \onlinecite{Hellman2011}, we observe a striking difference in harmonic  and TDEP force constants. This is illustrated in \figref{fig:frobnormar}. The effective TDEP force constants decrease faster with distance compared to the harmonic ones, a behavior that is expected. It is a vivid illustration of the temperature dependence of the potential energy landscape, and at the same time a confirmation that the TDEP technique describes this renormalisation well. From the free energy surface we can extract the finite temperature equation of state for bcc Zr, as illustrated in \figref{fig:zr_eos}.

\begin{figure}[htb]
\includegraphics[width=\linewidth]{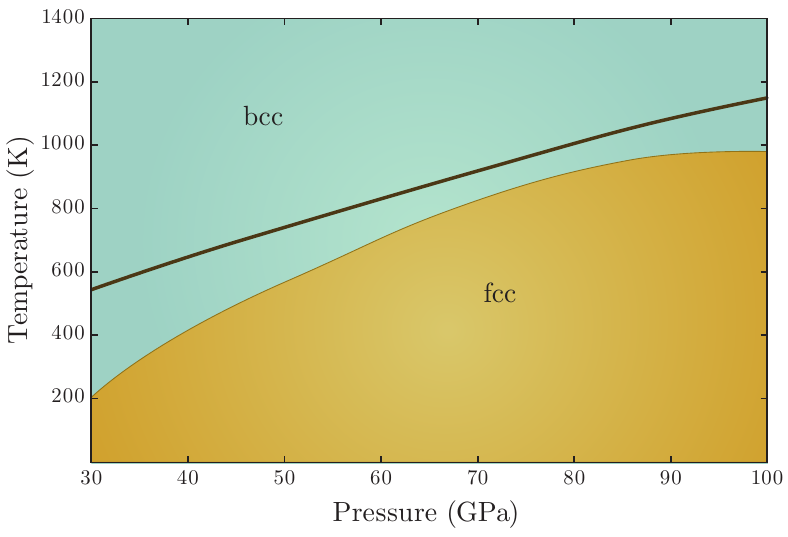}
\caption{\label{fig:he_phasediagram}(color online) The calculated phase diagram for $^4$He modelled with the Aziz \emph{et al.} potential.\cite{Aziz1979} The red line indicates the experimental melting curve. The observation of the stabilization of the bcc phase before the melting demonstrates that TDEP treats this system accurately and in agreement with other approaches even in such a strongly anharmonic system as $^4$He.}
\end{figure}

To test the performance of TDEP close to melting, we turn to solid He modelled with the Aziz \emph{et al.} potential\cite{Aziz1979}$^,$\footnote{$^4$He was modelled using the Aziz \emph{et al.} potential.\cite{Aziz1979} The bcc and fcc phases were both described by $5\times5\times5$ supercells (125 atoms), a 0.5fs timestep for a total simulation length of 20ps. Free energies were converged within 1meV/atom after about 5ps. Free energy surfaces were interpolated from a grid of six volumes and five temperatures in the appropriate range.}. The melting curve and bcc-fcc transition just before melting has been extensively studied, see Ref.~\onlinecite{Belonoshko2012} and references therein. As demonstrated in \figref{fig:he_phasediagram} strongly anharmonic He pose no problem for the presented method. The stabilization of the bcc phase before melting is consistent with results from phase-coexistence simulations. This are at the moment considered the most accurate methods for determining phase stabilities at high temperature. They do, however, require simulation cells much larger than what is accessible to AIMD, and can only be used with classical potentials. We show here that with the presented method we can with simulation sizes of 125 atoms accurately reproduce the same transition temperatures. This verifies the accuracy of the method and opens up the applicability to high pressure high temperature studies of phase stabilities close to melting. 

\section{Conclusions}

We have presented detailed description of the Temperature Dependent Effective Potential method for the treatment of lattice dynamics of strongly anharmonic solids, including an extension and refinement to this accurate technique. Moreover, we have detailed how the temperature dependence of all components of the free energy should be taken into account, and presented several successful examples, including a model anharmonic potential, first-principles calculations of equation of state for bcc Zr, and classical molecular-dynamics simulations of bcc-to-fcc transition in $^4$He.  

\section{Acknowledgements}

Support from the Knut \& Alice Wallenberg Foundation (KAW) project “Isotopic Control for Ultimate Material Properties”, the Swedish Research Council (VR) projects 621-2011-4426 and LiLi-NFM,  and the Swedish Foundation for Strategic Research (SSF) program SRL10-0026  is gratefully acknowledged. SSI acknowledges support from the Swedish Government Strategic Research Area Grant in Materials Science to AFM research environment at LiU. Supercomputer resources were provided by the Swedish National Infrastructure for Computing (SNIC). We would also like to thank Prof. Anatoly Belonoshko for suggesting the investigation of He phase transitions.


%

\end{document}